\documentclass[twocolumn]{aastex63}
\usepackage{threeparttable}
\usepackage{enumitem}
\usepackage{amsmath}
\usepackage{soul}
\usepackage{todonotes}

\newcommand{\degree}{\ensuremath{^\circ}}

\newcommand{\dmu}{pc~cm$^{-3}$}
\newcommand{\FIG}[1] {Figure~\ref{#1}}

\newcommand{\APP}[1] {Appendix~\ref{#1}}
\newcommand{\D}{\mathrm{d}}

\def\dme{\rm DM_E}

\def\ergs{\mathrm{erg}\, \mathrm{s}^{-1}}
\def\be{\begin{equation}}
\def\ee{\end{equation}}

\accepted{February 18, 2021}
\submitjournal{ApJL}
\shorttitle{CRAFTS for Fast Radio Bursts}
\shortauthors{Niu et al.}

\begin{document}

\title{CRAFTS for Fast Radio Bursts \\Extending the dispersion-fluence relation with new FRBs detected by FAST}

\correspondingauthor{Di Li} \email{dili@nao.cas.cn}

\correspondingauthor{Wei-Wei Zhu} \email{zhuww@nao.cas.cn}

\correspondingauthor{Chenhui Niu} \email{peterniu@nao.cas.cn}

\author[0000-0001-6651-7799]{Chen-Hui Niu}
\affiliation{National Astronomical Observatories, Chinese Academy of Sciences, Beijing 100101, China}

\author[0000-0003-3010-7661]{Di Li}
\affiliation{National Astronomical Observatories, Chinese Academy of Sciences, Beijing 100101, China}%
\affiliation{University of Chinese Academy of Sciences, Beijing 100049, China}
\affiliation{NAOC-UKZN Computational Astrophysics Centre, University of KwaZulu-Natal, Durban 4000, South Africa}

\author[0000-0002-4300-121X]{Rui Luo}
\affiliation{CSIRO Astronomy and Space Science, Australia Telescope National Facility, Epping, NSW 1710, Australia}

\author[0000-0001-9036-8543]{Wei-Yang Wang}
\affiliation{National Astronomical Observatories, Chinese Academy of Sciences, Beijing 100101, China}
\affiliation{School of Physics and State Key Laboratory of Nuclear Physics and Technology, Peking University, Beijing 100871,China}

\author[0000-0002-4997-045X]{Jumei Yao}
\affiliation{National Astronomical Observatories, Chinese Academy of Sciences, Beijing 100101, China}

\author[0000-0002-9725-2524]{Bing Zhang}
\affiliation{Department of Physics and Astronomy, University of Nevada, Las Vegas, Las Vegas, NV 89154, USA}

\author[0000-0001-5105-4058]{Wei-Wei Zhu}
\affiliation{National Astronomical Observatories, Chinese Academy of Sciences, Beijing 100101, China}

\author{Pei Wang}
\affiliation{National Astronomical Observatories, Chinese Academy of Sciences, Beijing 100101, China}

\author[0000-0002-3586-0786]{Haoyang Ye}
\affiliation{School of Astronomy and Space Science, Nanjing University, Nanjing 210093, China}

\author{Yong-Kun Zhang}
\affiliation{University of Chinese Academy of Sciences, Beijing 100049, China}

\author{Jia-rui Niu}
\affiliation{University of Chinese Academy of Sciences, Beijing 100049, China}

\author{Ning-yu Tang}
\affiliation{National Astronomical Observatories, Chinese Academy of Sciences, Beijing 100101, China}

\author{Ran Duan}
\affiliation{National Astronomical Observatories, Chinese Academy of Sciences, Beijing 100101, China}

\author{Marko Krco}
\affiliation{National Astronomical Observatories, Chinese Academy of Sciences, Beijing 100101, China}

\author{Shi Dai}
\affiliation{CSIRO Astronomy and Space Science, Australia Telescope National Facility, Epping, NSW 1710, Australia}
\affiliation{Western Sydney University, Locked Bag 1797, Penrith South DC, NSW 1797, Australia}

\author{Yi Feng}
\affiliation{University of Chinese Academy of Sciences, Beijing 100049, China}

\author{Chenchen Miao}
\affiliation{University of Chinese Academy of Sciences, Beijing 100049, China}

\author{Zhichen Pan}
\affiliation{National Astronomical Observatories, Chinese Academy of Sciences, Beijing 100101, China}

\author{Lei Qian}
\affiliation{National Astronomical Observatories, Chinese Academy of Sciences, Beijing 100101, China}

\author{Mengyao Xue}
\affiliation{National Astronomical Observatories, Chinese Academy of Sciences, Beijing 100101, China}

\author{Mao Yuan}
\affiliation{University of Chinese Academy of Sciences, Beijing 100049, China}

\author{Youling Yue}
\affiliation{National Astronomical Observatories, Chinese Academy of Sciences, Beijing 100101, China}

\author{Lei Zhang}
\affiliation{National Astronomical Observatories, Chinese Academy of Sciences, Beijing 100101, China}
\affiliation{School of Physics and Technology, Wuhan University, Wuhan 430072, China}

\author{Xinxin Zhang}
\affiliation{National Astronomical Observatories, Chinese Academy of Sciences, Beijing 100101, China}

\begin{abstract}
We report three new FRBs discovered by the Five-hundred-meter Aperture Spherical radio Telescope (FAST), namely FRB 181017.J0036+11, FRB 181118 and FRB 181130, through the Commensal Radio Astronomy FAST Survey (CRAFTS). Together with FRB 181123 that was reported earlier, all four FAST-discovered FRBs share the same characteristics of low fluence ($\leq$0.2~Jy~ms) and high dispersion measure (DM, $>1000$ \dmu), consistent with the anti-correlation between DM and fluence of the entire FRB population.
FRB 181118 and FRB 181130 exhibit band-limited features. FRB 181130 is prominently scattered ($\tau_s\simeq8$~ms) at 1.25~GHz. FRB 181017.J0036+11 has full-bandwidth emission with a fluence of 0.042~Jy~ms, which is one of the faintest FRB sources detected so far. CRAFTS starts to built a new sample of FRBs that fills the region for more distant and fainter FRBs
in the fluence-$\dme$ diagram, previously out of reach of other surveys. The implied all sky event rate of FRBs is $1.24^{+1.94}_{-0.90} \times 10^5$~sky$^{-1}$~day$^{-1}$ at the $95\%$ confidence interval above 0.0146~Jy~ms. We also demonstrate here that the probability density function of CRAFTS FRB detections is sensitive to the assumed intrinsic FRB luminosity function and cosmological evolution, which may be further constrained with more discoveries. 

\end{abstract}

\keywords{Radio transient sources (2008); Radio astronomy (1338); Astronomical object identification (87); Radio bursts (1339)}

\section{Introduction}
Fast radio bursts (FRBs) are bright, millisecond-duration cosmological radio transients (see \citealt{2019ARA&A..57..417C,2019A&ARv..27....4P,2020Natur.587...45Z} for reviews). The origin of FRBs remains a mystery (see \citealt{2019PhR...821....1P} for a summary of models). Based on their large dispersion measures (DMs) in excess of the expected corresponding Galactic contribution, their cosmological origin was first hypothesized and later confirmed \citep{2007Sci...318..777L,2013Sci...341...53T, 2017ApJ...843L...8B,2017Natur.541...58C,2017ApJ...834L...8M,2017ApJ...834L...7T,2019Natur.572..352R,2020Natur.577..190M,Macquart+20Nat}. 

Motivated by the ``Lorimer burst'' \citep{2007Sci...318..777L} and the confirmation of FRBs being a distinctive transient population \citep{2013Sci...341...53T}, a number of dedicated FRB surveys and searches have resulted in an increasing pace of discoveries (e.g., \citealt{2014ApJ...790..101S,2015Natur.528..523M,2017MNRAS.468.3746C,2018MNRAS.478.1209F,2019Natur.572..352R,2019Natur.566..230C,2019MNRAS.488.2989F}), with the anticipation of the upcoming new CHIME sample being a substantial leap forward.

The last sentence in the abstract of \citet{2013Sci...341...53T}-- ``Characterization of the source population and identification of host galaxies offers an opportunity to determine the baryonic content of the universe."--turns out to be both insightful and visionary. There exists an anti-correlation between DM and fluence of FRBs, which is consistent with DM being a proxy of distance \citep{Shannon+18Nat}. The relation has a broad scatter, implying a broad range of intrinsic luminosity and/or energy functions of FRBs (see also \citealt{Luo+18MN,Luo+20MN,2020MNRAS.498.1397L,2020MNRAS.498.1973L,2021MNRAS.501..157Z}). 
The observed sample is obviously limited by the available instruments. In addition to event rate and other intrinsic characteristics of FRBs, the instrumental gain, RFI, and detection algorithms all affect the apparent fluence distribution. Still, high-DM FRBs are most likely luminous and distant, thus crucial for extending the dispersion-fluence relation. For example, the source with the highest DM $\sim2600~\rm pc~cm^{-3}$ \citep{2018MNRAS.475.1427B,2018MNRAS.478.2046C} detected thus far, FRB 160102, can be inferred to have an upper-limit of redshift $z\sim3$ \citep{Zhang18ApJL}.

The Five-hundred-meter Aperture Spherical radio Telescope (FAST,\citealt{2011IJMPD..20..989N,li16,2019RAA....19...16L}), while non-competitive in terms of FRB discovery rate, is well suited for catching distant bursts. The brightest known FRB pulses can be detected up to $z\sim10$ \citep{Zhang18ApJL}, if they exist there. Based on a novel high-cadence CAL technique, the Commensal Radio Astronomy FAST Survey (\footnote{\url{https://crafts.bao.ac.cn}} \citealt{8331324}) aspires to cover the FAST sky ($\sim 57\%$ of the full sky) with the 19-beam receiver in drift scan mode and simultaneously obtain four data streams, namely Galactic HI, extra-galactic HI, pulsar search, and transients, the last two of which are searched for FRBs. As a drift scan, CRAFTS spends most of its time off the Galactic plane($85\%$ of observation time with $|b|>10\degree$). 
In \citep{Zhu+20ApJ} the first FAST discovery--FRB181123--was reported. In this paper, we report three new FRBs. Together, the four discoveries, with high-DM as well as low fluence, occupy a previously empty region ($\rm DM_E \in [ 1099.4 pc~cm^{-3} , 1819.3~pc~cm^{-3} ] $ , Fluence $\in [ 42 \rm ~mJy~ms,200~mJy~ms ] $) in the DM-fluence space. 

We describe the observation and searching methods in Sec.\ref{sec:Observation and data processiong} and the characteristics of the three new FRBs in Sec.\ref{sec:3frbs}. We discuss the all sky FRB event rate expected for CRAFTS, the implications of these High-DM and low-fluence FRBs in Sec.\ref{sec:analysis}. The summary is in Sec.\ref{sec:discuss}.

\section{observations}\label{sec:Observation and data processiong}
The FAST's L-band receiver Array of 19-beams (FLAN, \citealt{8331324}) covers a frequency band of 1.0-1.5~GHz.
CRAFTS employs a novel high-cadence CAL injection technique invented by the survey team, which produced a system temperature measurement for each detected pulse. In this work, we search for FRBs in the pulsar data stream, with 0.122~MHz frequency resolution at 196.608~$\mu$s sampling interval. The raw data are of 8-bits sampling with full polarization, reaching data rate of $\sim 300$~GB per hour. The observed data were compressed to the 1-bit filterbank format and the polarizations were merged. 

We established a \texttt{HEIMDALL}\footnote{\url{http://sourceforge.net/projects/heimdall-astro} \citep{2012MNRAS.422..379B}}-based pipeline called \texttt{FAST$\_$Miner}\footnote{\url{https://github.com/peterniuzai/FAST_FRB.git}}.
\texttt{FAST$\_$Miner} distributes the workload to more than 20~GPU servers. The pipeline records the S/N (Signal-to-Noise ratio), start time, pulse width, DM, beam ID, etc. 
For each single-pulse candidate, the pipeline produces an overview plot including two profiles and two- dimensional waterfall plots, corresponding to before and after the de-dispersion process, respectively.

For a total of 1667 hours CRAFTS data taken between June and the end of 2018, \texttt{FAST$\_$Miner} were run with a DM trial range between 100 to 5000~pc~cm$^{-3}$. Candidates detected by less than 4 adjacent beams but with a S/N $> 7\sigma$ were kept for further inspection.The candidates were then manually examined, particularly in terms of the DM sweep in the dynamic spectrum and their surrounding RFI background. According to the radiometer equation:
\begin{equation}\label{Eq:radiometer equation}
 S_{\rm peak} = \frac{\beta \cdot T_{\rm sys}\cdot \rm S/N} {G\sqrt{2~\Delta t~\rm \Delta\nu}} \>,
\end{equation}
where $\beta$ is $\pi/2$ for the 1-bit digital quantization and $\Delta\nu$ is the bandwidth, the detection threshold of our CRAFTS FRB search is roughly 0.0146~Jy~ms for a 1~ms pulse width.

\section{FRB Discoveries}\label{sec:3frbs}
\begin{table*}[htbp]

\begin{center}
\caption{Properties of three new FRBs detected in CRAFTS}
\begin{threeparttable}

\begin{tabular}{lccc}

\hline \hline

FRB YYMMDD(.J2000) & FRB~181017.J0036+11 & FRB~181118 & FRB~181130 \\
\hline
 \multicolumn{4}{c}{\textbf{Measured Parameters}}\\
 
Event MJD at 1.5~GHz              & 58408.665197 & 58440.877654 & 58452.542674 \\
FAST Beam ID (M01 -- M19)               & M14  & M07  & M11 \\
Right Ascension (J2000)           & 00$^\mathrm{h}$36$^\mathrm{m}$29$^\mathrm{s}$.8 & 
                                    07$^\mathrm{h}$56$^\mathrm{m}$41$^\mathrm{s}$.87 & 
                                    00$^\mathrm{h}$39$^\mathrm{m}$07$^\mathrm{s}$.85 \\
Declination (J2000)               &  11$^\circ$19$^\prime$59$^{\prime\prime}$.8 & 
                                    16$^\circ$08$^\prime$56$^{\prime\prime}$.7 &
                                    19$^\circ$24$^\prime$31$^{\prime\prime}$.7 \\
Galactic Coordinates ($l,b$)      & $117^\circ.9,-51^\circ.4$ & $205^\circ.2,21^\circ.5$  & $118^\circ.9,-43^\circ.4$ \\
Dispersion Measure (\dmu)         & 1845.2$\pm1$ & 1187.7$\pm3.3$ & 1705.5$\pm6.5$\\
Emission Freq. (GHz)              & 1.0 -- 1.5 & 1.00 -- 1.12 & 1.16 -- 1.50 \\
Dispersion smear  (ms)\tnote{a}   & 1.0  & 1.0 & 0.7 \\
Measured width (ms)\tnote{b}      & 1.43$\pm0.25$ & 5.3$\pm3
.72$ &  $9.52^{+5.94}_{-5.08}$\\
Scattering effect at 1.25GHz (ms) & $<0.5$  & $<2.08$ &  $7.64^{+4.96}_{-4.51}$ \\
Measured S/N                      & 14.3 & 11.0  & 30.9 \\
Observed peak flux density (mJy)  & $\sim34.1$    & $\sim14.2$   & $\sim20.6$ \\
Measured fluence (Jy ms)\tnote{c}          & 0.042   & 0.064  & 0.168  \\

\multicolumn{4}{c}{\textbf{Inferred Parameters}\tnote{d}}\\

$\rm DM_{Gal}$ (\dmu)\tnote{e}    &  34.62 , 25.91 & 71.53 , 88.30 & 38.16 , 29.69 \\
Estimate redshift ($z$)\tnote{f}  &  $2.01^{+0.04}_{-0.04}$ , $2.02^{+0.04}_{-0.04}$   &
                                     $1.17^{+0.05}_{-0.06}$ , $1.15^{+0.05}_{-0.06}$   &
                                     $1.83^{+0.04}_{-0.05}$ , $1.84^{+0.04}_{-0.04}$  \\
Max. comoving distance (Gpc)      &  5.4   & 4.0   & 5.1 \\
Max. luminosity distance (Gpc)    &  16.5  & 8.7   & 14.8\\
Max. isotropic energy ($10^{40}\,\rm erg$)\tnote{g} &  0.45   & 0.26 & 1.5 \\
Average luminosity ($10^{43}\,\ergs$) & 0.3 & 0.05 &  0.16\\
\hline \hline
\end{tabular}
\label{tab:3FRBs}
\begin{tablenotes}

\footnotesize
    \item[a] The referenced frequency for smear calculation takes center of the emission frequency band.
    \item[b] Full pulse width at the half maximum from Gaussian fitted profile. Note that the intrinsic width can be much narrower for FRB181017 since we do not save the raw data \citep{2020MNRAS.497.1382Q}.
    \item[c] Take the gain at the center of the beam.
    \item[d] The parameters for Cosmology model are from \cite{2020A&A...641A...6P}.
    \item[e] $\rm DM_{Gal}$ denotes the DM contribution from the Galaxy, and is calculated using the NE2001 and the YMW16 model, respectively.
    \item[f] Redshifts inferred from the extra-galactic DM calculated using the NE2001 and the YMW16 model, respectively. The corresponding deduction and error analysis can be found in \APP{app:redshift estimation}.
    \item[g] Calculated by assuming a flat spectrum within a width of 1~GHz tentatively, due to a lack of complete spectrum measurement for FRBs currently.
\end{tablenotes}

\end{threeparttable}
\end{center}
\end{table*}

Three new FRBs were captured on October 17th, November 18th, and November 30th in 2018. Since two other distinctive FRBs have already been reported to occur on October 17th \citep{2019ApJ...885L..24C,2019MNRAS.488.2989F}, we denote the CRAFTS discovery as FRB181017.J0036+11 to avoid confusion. All the detailed results of the three new FRBs are exhibited in Table \ref{tab:3FRBs}.

\subsection{Burst Characteristics}
FRB 181017.J0036+11, FRB 181118, and FRB 181130 were detected within different FLAN beams. We used the average gain ($G \sim 15$ K/Jy, \citealt{2020RAA....20...64J}) with Zenith angle correction and the average system temperature ($T_{\rm sys}\sim$ 20~K) to estimate the peak fluxes. The 0.042~Jy~ms fluence of FRB 181017.J0036+11 is the lowest one-off event ever reported in the FRB Catalog \citep{2016PASA...33...45P}. 

\begin{figure*}[htbp]
\centering
\includegraphics[width=1\textwidth]{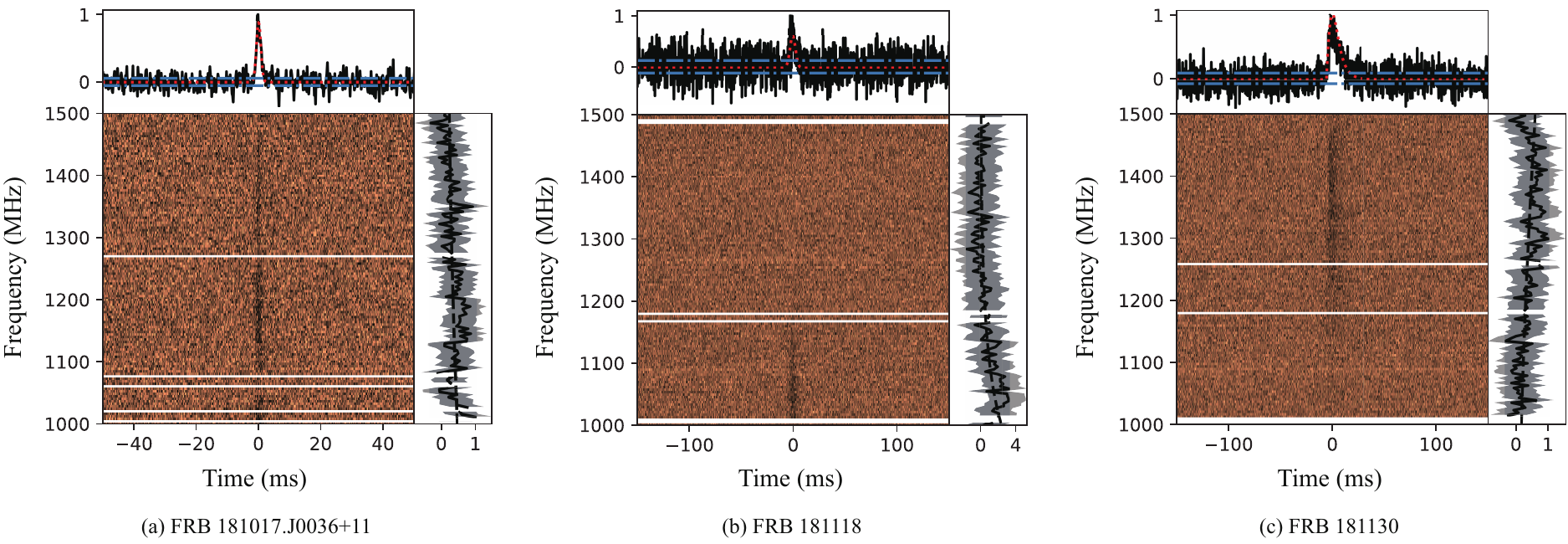}
\caption{The center plot of (a) (b) and (c) shows the 2-dimensional dynamic spectrum of the three newly-discovered FRBs where RFIs are masked. The top panel of each plot is the normalized pulse profile, while the right panel shows the characteristic spectrum within 2$\sigma$ range(gray shadow) of each burst. No scattering tail was found in the profiles of either FRB 181017 (a) or FRB 181118 (b) at their intrinsic DM values, so Gaussian fittings are outlined. FRB 181130 exhibits an evident scattering effect on a Gaussian-fitted profile.}\label{fig:waterfall}
\end{figure*}

The pulse profiles and waterfall plots of the three new FRBs are presented in \FIG{fig:waterfall}. FRB 181017.J0036+11 swept across the full observing band of 500~MHz, whereas FRB 181118 and FRB 181130 are band-limited. Both FRB 181017.J0036+11 and FRB 181118 show a decrease in the power as the frequency increases, which is similar to the negative spectra index commonly seen in pulsars, while FRB 181130 behaves in the opposite way. The available observing bandwidth and SNR is insufficient for a direct determination of the FRB spectral index.

\subsection{Propagation effects}
The $\rm DM_{obs}$ for each FRB is estimated through \citep{2019ApJ...876L..23H}
\begin{equation}\label{Eq:Multi_peak}
 \rm{DM_{obs}} = Max\left\lbrace \sum\left(\dfrac{ d\rm{I}(\rm{DM},t)}{d t}\right)^2\right\rbrace,
\end{equation}
where $\rm{I}(\rm{DM},t)$ stands for the burst profile at a given DM and sampling time $t$. 
All three FAST FRBs exhibit high DMs ($\rm DM_{obs}$ in Table~\ref{tab:3FRBs}).
The highest $\rm DM_{obs}$ in this work is $\rm DM_{obs}$=1845.2 \dmu\ for FRB 181017.J0036+11. The Full Width at Half Maximum (FWHM) of the fitted DM is taken as the uncertainty.
Considering the high Galactic latitudes of these events, these FRBs are most likely to have cosmological origins, which we analyze in \APP{app:redshift estimation}.

Following \cite{mls+15}, we were able to obtain a scattering timescale $7.64^{+4.96}_{-4.51}$~ms for FRB 181130 as shown in panel (c) of \FIG{fig:waterfall},the details are in \APP{app:scatter analysis}.
No measurable scattering was seen for the other two sources, but the upper limits are given according to the sharpest decline in signal power with time. No measurable scintillation bandwidth was found for all three.

\subsection{Follow-up Observations}
On April 4th and 5th, 2020, we tracked all three new FRBs for one-hour each source with FLAN. 
We tracked FRB 181123 for a total of 6 hours between February and July of 2020.
The data with higher time resolution (98~$\mu$s) with full stokes polarization were recorded. No more burst was detected in any of the follow-up observation.

\section{Implications and Discussion}\label{sec:analysis}
We look into the implications of FAST discoveries for the event rate, DM-fluence relation, and the intrinsic luminosity of FRBs.

For a Poissonion FRB distribution in an isotropic universe and an effective beam area of 0.019 d$^2$ and taking the events possibility from \cite{1986ApJ...303..336G}, four FRBs in 1667 hours drift scan and 9 hours follow-up observations corresponds to an all sky event rate as $1.24^{+1.94}_{-0.90} \times 10^5$~sky$^{-1}$~day$^{-1}$ at the $95\%$ confidence level above 0.0146~Jy~ms (7$\sigma$ for a 1~ms duration), which is the lowest fluence threshold ever considered.

\begin{figure}[htbp]
\centering
\includegraphics[width=0.45\textwidth]{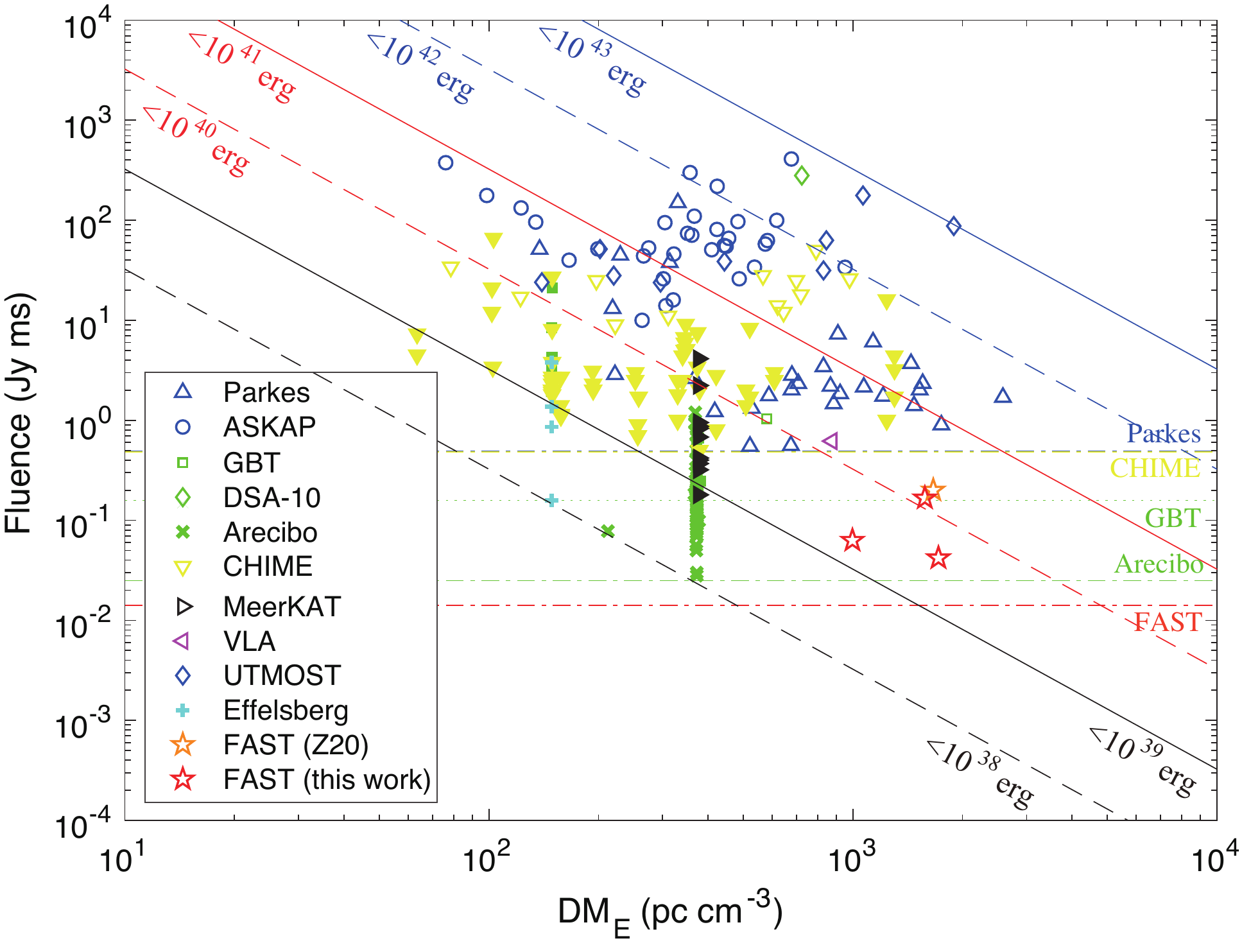}
\caption{\small{A radio fluence-$\rm DM_E$ diagram for FRBs, where $\rm DM_E$ represents the DM value after deducting the DM contribution from the Milky Way. With the assumption of an isotropic beam and $\rm DM_{host}=0$, contours of the energy upper limits are plotted in slanted lines, with respective numerical values of $E$ labeled. The sensitivities of the telescopes are marked by horizontal lines. Note that the FAST sensitivity is for the CRAFTS mode (1-bit quantization in 2018 data) and will be even lower for other modes.
The FRBs detected by different telescopes are represented by different symbols. In particular, orange pentagram represents the FRB 181123 in \cite[][ Z20]{Zhu+20ApJ} and the red pentagrams are FRBs reported in this work. The filled symbols represent known repeating FRBs.} }
\label{fig:fluence and DM}
\end{figure}

\FIG{fig:fluence and DM} shows the updated fluence-$\dme$ relation with the FAST events added, where $\rm DM_E$ represents the DM value after deducting the DM contribution from the Milky Way \citep{2002astro.ph..7156C,2017ApJ...835...29Y}. Together with FRB 181123 reported in \cite{Zhu+20ApJ}, the four FAST FRBs clusters to the bottom right of \FIG{fig:fluence and DM}, which are consistent with the extrapolation of ASKAP FRB fluence to higher $\rm DM_{E}$ \citep{Shannon+18Nat}.
The DM-fluence relation has been extended to the more distant and low fluence region in the DM-fluence space.

Identifying the host galaxies of these FRBs and measuring their redshifts will help extend the ${\rm DM}-z$ relation (also known as the Macquart relation) \citep{Macquart+20Nat,liz20} to higher redshifts. Alternatively, a low fluence burst could be due to a low luminosity of a nearby FRB. In this case, the large DM must be associated with the nearby environment or host galaxy of the FRB. Identifying these low-luminosity events would be essential to constrain the low end of FRB luminosity function \citep{Luo+18MN}. 
Given the short integration realized by drift scan, systematic long integration on apparent non-repeaters is necessary to distinguish whether repeaters have low intrinsic luminosity or the we just under-sampled the non-repeaters.
This points to the value of systematic long integration on apparent non-repeaters to distinguish effects of under-sampling vs lower intrinsic luminosity.

\begin{figure}[htbp!]
    \centering
    \includegraphics[width=0.45\textwidth]{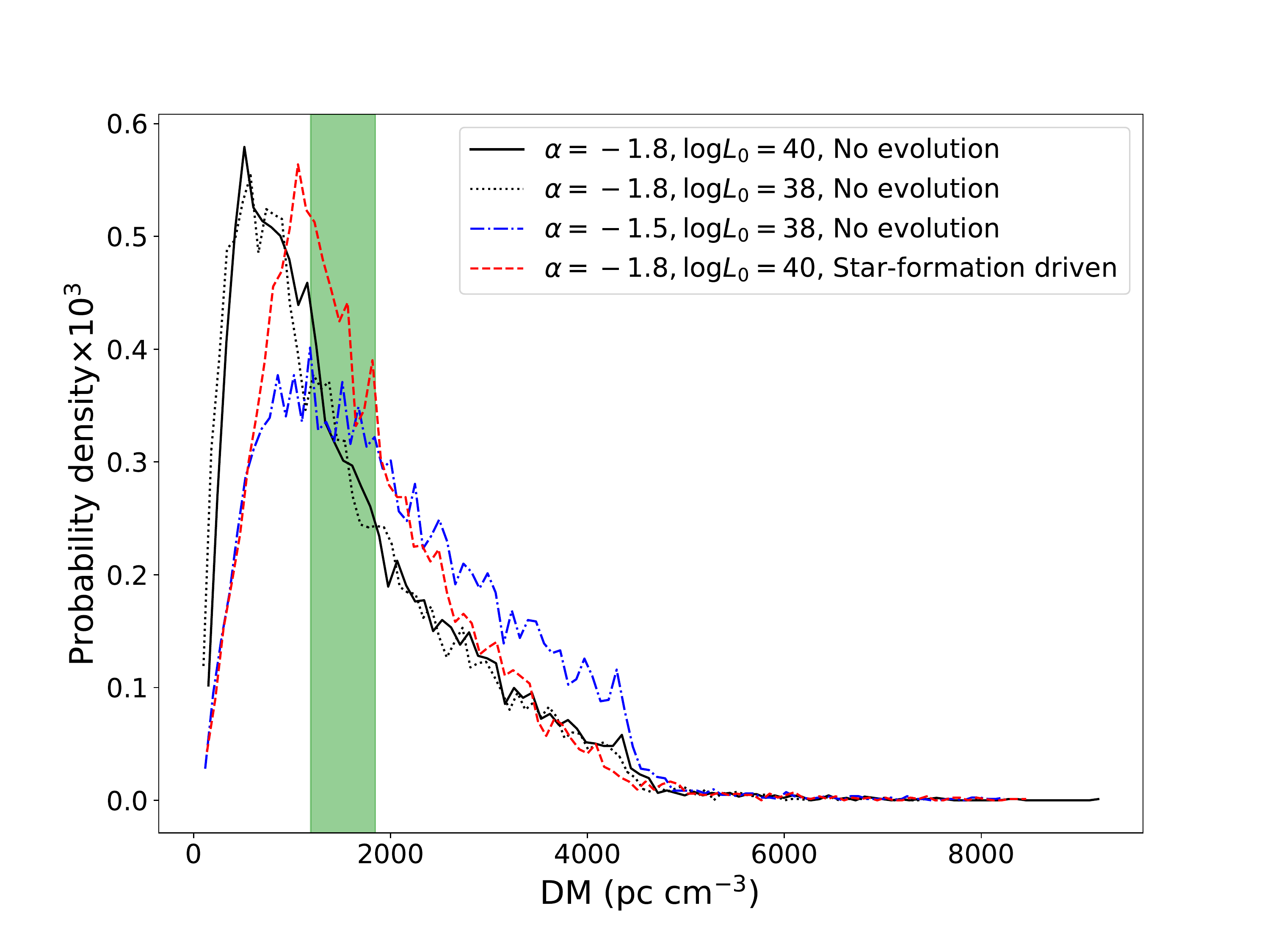}
    \caption{The PDF of the DM distribution of FRBs would be detectable by the FAST telescope. The x-axis is the DM value in units of \dmu, and the y-axis is the probability density scaled up by a factor of 1000. The four curves represent the DM PDFs of FRBs detectable by the FAST in our four simulation cases: $\alpha=-1.8$ and $L_0=10^{40}\,\ergs$ without cosmological evolution (black solid), $\alpha=-1.8$ and $L_0=10^{38}\,\ergs$ without cosmological evolution (black dotted), $\alpha=-1.5$ and $L_0=10^{38}\,\ergs$ without cosmological evolution (blue dashed-dotted), and $\alpha=-1.8$ and $L_0=10^{40}\,\ergs$ with star-formation history as cosmological evolution (red dashed).} The green shaded rectangle region shows the DM range of the four FRBs detected so far in CRAFTS.
    \label{fig:dm_pdf}
\end{figure}

Following \cite{Luo+18MN} and \cite{Luo+20MN}, we quantify the DM distribution of FAST detections by assuming a Schechter luminosity function for all FRBs
\begin{equation}
\phi(L)\,\D L=\phi^* 
\left(\frac{L}{L^*}\right)^{\alpha}e^{-\frac{L}{L^*}}\,\D \left(\frac{L}{L^*}\right)\, ,
\label{eq:lf}
\end{equation}
where $\alpha$ is the power-law index, and $L^*$ is the upper cut-off luminosity. The lower cut-off luminosity is defined as the intrinsic minimum luminosity of the FRB population, which was denoted as $L_0$. We then carried out mock observation of the FRBs generated based on Eq.~\ref{eq:lf} as well as under an assumption of cosmological evolution (e.g. following the star-formation history).
The specifics are described in detail in \APP{app:simu_dm}. The probability density function (PDF) of DM expected for the FAST sample thus simulated is shown in \FIG{fig:dm_pdf}.

Three points to note from our simulation. First, the PDF of FAST-FRB's DM is more sensitive to $\alpha$ than $L_0$. Second, the non-evolving flatter luminosity function ($\alpha=-1.5$) provide the best apparent match to the observation, although including evolution will also move the PDF peak to higher DM. 
Third, for all the model parameters explored here, FAST will have significant detection probability ($>$10\%) for $\mathrm{DM}>3000$~pc~cm$^{-3}$, which is beyond reach for other surveys.

\section{Summary}\label{sec:discuss}
In this paper, we report three new highly dispersed and low-luminosity FRBs discovered by the Commensal Radio Astronomy FAST Survey, which is a multi-purpose drift scan survey with the 19-beam system of FAST. Our main findings are as follows.

1) Along with the first discovery--FRB181123--previous reported, four events in a total of $\sim$1676 hours, corresponds to an all sky rate of $1.24^{+1.94}_{-0.90} \times 10^5$~sky$^{-1}$~day$^{-1}$ at the $95\%$ confidence interval above 0.0146~Jy~ms, by far the deepest such estimate.

2) All three new FRBs are off the Galactic plan, with Galactic latitudes $|b|=21.5^\circ$, $43.4^\circ$, and $51.4^\circ$, for FRB 181118, FRB 181130, and FRB 181017.J0036+11, respectively. The estimated Galactic contribution $\rm DM_{MM}$ are all less then 10\% of the total DM, suggesting extra-galactic origins.

3) The measured DM, 1187.7$\pm3.3$, 1705.5$\pm6.5$, and 1845.2$\pm1$, when with $\rm DM_{\rm MW}$ subtracted, correspond to upper limits of the estimated redshifts of these events between $z=1.17$ and $z=2.01$.

4) FRB 181017.J0036+11, with a pulse width of 1.43~ms and 34.1~mJy peak flux, has the lowest fluence (0.042~Jy~ms) among all the non-repeaters.

5) A prominent scattering tail in the profile of FRB 181130 is found to be consistent with a 7.64~ms scattering time scale. No measurable scattering or scintillation effect can be found for other sources.

6)We carried out an FRB population synthesis study. The CRAFTS sample, given its high median DM and low fluence, are consistent with a Schechter-type intrinsic luminosity function with the DM distribution being sensitive to both the slope $\alpha$ and cosmological evolution.

7) These CRAFTS FRBs occupies a previously empty region in the DM-fluence space. With more discoveries and possible localization, they have the potential to extend the DM-$z$ Macquart relation.

\section*{acknowledgments}
This work is supported by National Key R$\&$D Program of China No.2017YFA0402600 and is partially supported by the National Natural Science Foundation of China Grant No.\ 11988101,11725313, 12041303, 11873067,the Cultivation Project for FAST Scientific Payoff and Research Achievement of CAMS-CAS, the CAS-MPG LEGACY project, the Strategic Priority Research Program of the Chinese Academy of Sciences Grant No. XDB23000000 and the National SKA Program of China No. 2020SKA0120200. C-H. Niu is grateful for the funding of the FAST Fellowship. 

\appendix

\section{Redshift estimation}\label{app:redshift estimation}
We describe the method to estimate redshifts of the three new FAST-discovered FRBs here.
The observed DM is generally given by
\begin{equation}
\rm DM_{obs}=DM_{MW}+DM_{halo}+DM_{IGM}+\frac{DM^{Loc}_{HG, sr}}{1+\it z},
\label{eq:DM}
\end{equation}
where $\rm DM_{MW}$ is the DM contribution from the Milky Way galaxy, $\rm DM_{halo}$ denotes the DM contribution associated with the Milky Way halo, $\rm DM_{IGM}$ represents the DM contributed by the IGM, and $\rm DM^{Loc}_{HG, sr}$ denotes DM contributed by the source itself and its host galaxy in the source frame, respectively.
Here, $\rm DM_{MW}$ can be derived from the Galactic electron density models \citep{2002astro.ph..7156C,2017ApJ...835...29Y}. 
The differences between results calculated using the two models for these FRBs are several tens $\rm pc\,cm^{-3}$.
The DM contribution associated with the Milky Way halo is assumed as $\rm DM_{halo}=30\pm15\,pc\,cm^{-3}$ following \cite{2015MNRAS.451.4277D}.
The average DM contributed by the IGM for $\Lambda$CDM cosmology is \citep{2014ApJ...783L..35D}
\begin{equation}
{\rm DM}_{\rm IGM} = \frac{3cH_0\Omega_bf_{\rm IGM}}{8\pi G m_p}\int^z_0\frac{\chi(z)(1+z)dz}{[\Omega_m(1+z)^3+\Omega_{\lambda}]^{\frac{1}{2}}},
\label{eq:IGM}
\end{equation}
where the free electron number per baryon in the universe is $\chi(z) \approx 7/8$ and the fraction of baryons $f_{\rm IGM}\sim0.83$.
Parameters of the cosmological model are taken following \citep{2020A&A...641A...6P}.
In order to derive the redshift $z$, we assume that the DM contribution from the host galaxy $\rm DM^{Loc}_{HG, sr}$ is in the range of 0 to 200 \dmu, which is consistent with several simulated results \citep{2015RAA....15.1629X, Luo+18MN, Luo+20MN} as well as results constrained from the observational data \citep{liz20}.
The inferred redshifts are listed in Table~\ref{tab:3FRBs}.

\section{The pulse width and scattering timescale of FRB 181130}\label{app:scatter analysis}
\begin{figure}[htbp]
    \centering
    \includegraphics[width=0.8\textwidth]{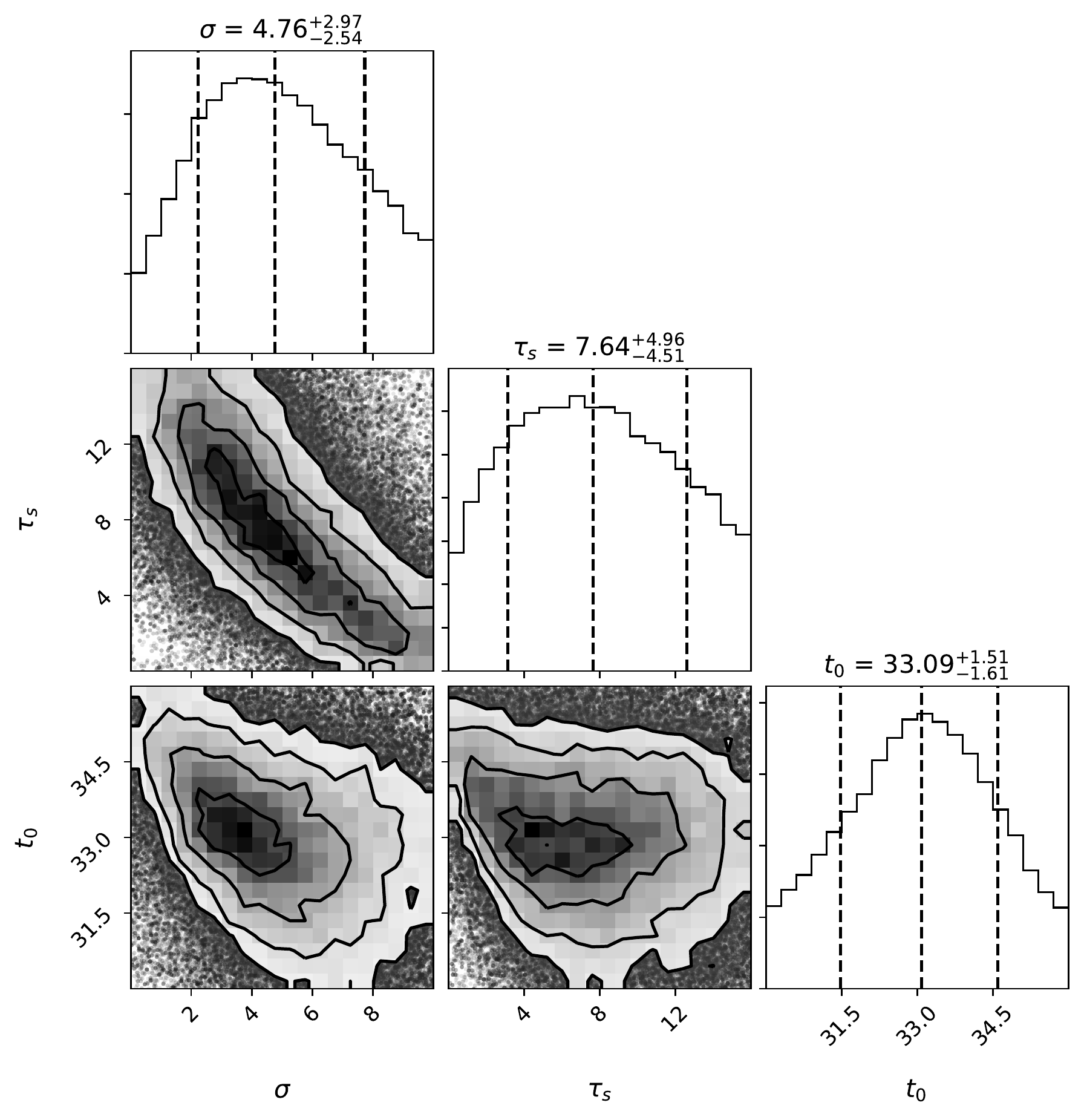}
    \caption{Posterior distributions of $\sigma$, $\tau_s$ (centered at 1250~MHz) and $t_0$ obtained by using \texttt{emcee}. The dashed lines represent 16, 50 and 84 percentiles in the corresponding histograms.}
    \label{fig:scatter_frb181130}
\end{figure}

To measure the pulse width and scattering timescale of FRB 181130, we first fold the frequency and normalized the pulse profile using the dynamic spectra in \FIG{fig:waterfall}. Following \citet{mls+15}, we then analyze the 78~ms-long data centered around the pulse peak by using \texttt{emcee} \citep{fhl+13}. In \FIG{fig:scatter_frb181130}, we show the data fitting results for the width $\sigma$ of the intrinsic Gaussian pulse component, the scattering timescale $\tau_s$ (centered at 1250~MHz) and the pulse arrival time $t_0$ relative to the starting point of the selected data. 

\section{Simulations on FRB detectability for CRAFTS}
\label{app:simu_dm}
Referring to the verification section for Bayesian inference in \cite{Luo+18MN, Luo+20MN}, we can also use the Monte-Carlo method to simulate FRBs that are detectable by FAST surveys such as CRAFTS. The specific procedures are described as follows:

\begin{enumerate}[label=(\roman*)]
    \item Sample the FRB luminosities $L$ according to the LF parameters measured in \cite{Luo+20MN}.
    \item Sample the host redshifts for two cases: 1) Non-evolved: $z$ merely follows a uniform distribution of FRBs across the universe; 2) Star-formation driven: $z$ follows both galaxy uniform distribution and the star-formation history described in \cite{2021MNRAS.501..157Z} \footnote{The current FRB sample is too small to constrain FRB redshift distribution models \citep{2019ApJ...883...40L, 2021MNRAS.501..157Z}, so we use the simplest $z$-distribution model in our simulations.}. Then calculate the DM values contributed from the intergalactic medium.
    \item Sample the intrinsic FRB pulse widths in the local rest frame of FRBs using the log-normal distribution constrained in \cite{Luo+20MN}.
    \item Based on the cosmologically evolving DM distributions of host galaxies at redshifts $z$ described by \cite{Luo+18MN}, generate the DM contributions from host galaxies in the local rest frame of FRBs. 
    \item Generate DM values contributed by the local circumstance of FRB progenitors using the uniform distribution assumed in \cite{Luo+18MN}.
    \item Generate Galactic DM values using the YMW16 model, and then sum the DMs from all of components mentioned above to obtain the total observed values.
    \item Produce a number of FRB positions randomly falling inside the FAST's beams, and then calculate the received peak flux density using the simulated luminosities, redshifts and beam responses of the FRB positions within the beam size.
    \item Based on the emitted pulse widths, redshifts and DMs of FRBs obtained in the steps above, calculate the observed pulse width considering DM smearing and cosmological time dilution.
    \item Select the FRBs whose peak fluxes are above the threshold of CRAFTS as positive detections.
\end{enumerate}

With the simulated data from mock FRBs, we then obtain the PDF of DMs for the CRAFTS' detectable FRBs, which 
are shown in \FIG{fig:dm_pdf}.


\end{document}